\documentclass[a4paper,11pt]{article}
\usepackage{pos}
\usepackage{caption}
\usepackage{subcaption}

\title{\texttt{nectarchain}, the scientific software for the Cherenkov Telescope Array - NectarCAM}
 \ShortTitle{\texttt{nectarchain}, the scientific software for the CTA-NectarCAM}
 
\author*[a]{G.~Grolleron}
\author[b]{H.~Ashkar}
\author[c]{F.~Brun}
\author[d]{H.~Costantini}
\author[h]{D.~Dumora}
\author[e]{P.~Jean}
\author[f]{D.~Kerszberg}
\author[a]{J.-P.~Lenain}
\author[c]{V.~Marandon}
\author[g]{S.~R.~Patel}
\author[e]{L.~Tibaldo}

\affiliation[a]{Sorbonne Université, CNRS/IN2P3, Laboratoire de Physique Nucléaire et de Hautes Energies, LPNHE, 4 place Jussieu, 75005 Paris, France}
\affiliation[b]{Laboratoire Leprince-Ringuet, École Polytechnique (UMR 7638, CNRS/IN2P3, Institut Polytechnique de Paris), 91128 Palaiseau, France}
\affiliation[c]{IRFU, CEA, Université Paris-Saclay, Bât 141, 91191 Gif-sur-Yvette, France}
\affiliation[d]{Aix-Marseille Université, CNRS/IN2P3, CPPM, 163 Avenue de Luminy, 13288 Marseille cedex 09, France}
\affiliation[e]{IRAP, Université de Toulouse, CNRS, CNES, UPS, 9 avenue Colonel Roche, 31028 Toulouse, Cedex 4, France}
\affiliation[f]{Institut de Fisica d'Altes Energies (IFAE), The Barcelona Institute of Science and Technology, Campus UAB, 08193 Bellaterra (Barcelona), Spain}
\affiliation[g]{Laboratoire de Physique des 2 infinis, Irene Joliot-Curie,IN2P3/CNRS, Université Paris-Saclay, Université de Paris, 15 rue Georges Clemenceau, 91406 Orsay, Cedex, France}
\affiliation[h]{Laboratoire de Physique des 2 infinis, Université de Bordeaux, CS 10120,19 chemin du Solarium, 3175 Gradignan, Cedex, France}

\onbehalf{for the CTA NectarCAM Project} 

\emailAdd{ggroller@lpnhe.in2p3.fr}

\abstract{The NectarCAM is a camera that will be mounted on the Medium-Sized Telescopes of the Cherenkov Telescope Array (CTA) observatory. Along with the hardware integration of the camera, the scientific software, \texttt{nectarchain}, is being developed. The software is responsible for transforming the raw data from the camera into analysis-ready calibrated data. In this contribution, we present the structure of the software, which consists of two modules: the calibration pipeline and the data quality check pipeline. The calibration pipeline reduces the data, performs flat fielding, and determines the gain for the analysis. The data quality monitoring pipeline is used to select the data that meets the necessary standards for analysis. Additionally, we discuss the format of the downstream data and the integration of the \texttt{nectarchain} modules in the general software framework of CTA. We also present the necessary tests for validating each part of the code. We conclude by mentioning the prospects for the future of the software.}

\ConferenceLogo{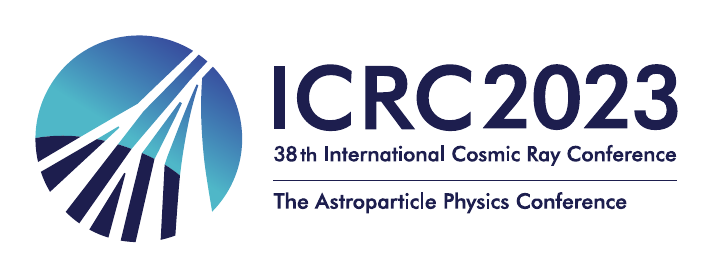}

\FullConference{%
38th International Cosmic Ray Conference (ICRC2023)\\
  26 July - 3 August, 2023\\
  Nagoya, Japan}

\begin{document}
\maketitle

\section[Introduction]{Introduction}
\label{Introduction}

The Cherenkov Telescope Array (CTA) will be the next generation ground-based imaging atmospheric Cherenkov telescopes (IACT) observatory. In comparison to the current IACT arrays, it will improve the sensitivity by a factor five to ten depending on the energy. Therefore it will bring a new outlook of the Universe. CTA will be composed of three types of telescopes, the Small-Sized-Telescopes (SST), the Medium-Sized-Telescopes (MST) and the Large-Sized-Telescopes (LST). They will be sensitive to the lower, median and higher part of the CTA energy range respectively from 20 GeV to beyond 100 TeV.

NectarCAM \citep{2017AIPC.1792h0009G} will be the camera mounted on the MSTs located in la Palma, the northern site of the Cherenkov Telescope Array Observatory (CTAO). This camera has a 8 degrees field of view and consists of 265 modules each equipped with 7 photomultiplier tubes (PMTs) able to detect Cherenkov light produced by very high energy (VHE) photons entering in the atmosphere. Each module is equipped with a trigger and readout electronic system, which operates with GHz sampling on 60 ns windows. The camera has two gain channels : the high gain to measure signals at the single photo-electron (p.e.) level and the low gain which aims at reconstructing signal up to 2000 p.e.. A full camera prototype is at the CEA Paris Saclay test bench where some tests have been performed (timing, temperature, etc.) and where flat-field (FF), pedestal (Ped) and gain calibration data acquisitions are taken to calibrate the camera and to develop the software which will be used on site to generate the calibrated data (R1) from the raw data produced by the camera (R0).

In this proceeding, we are focusing on the software pipeline \texttt{nectarchain}\footnote{\url{https://github.com/cta-observatory/nectarchain}}. In section \ref{Main Goal : R0 to R1 calibration}, the main goal of the calibration pipeline is presented. Then in section \ref{nectarchain modules}, the \texttt{nectarchain} workflow is presented and the current development status is explained. Next the interface with \texttt{ctapipe} in presented in section \ref{Interface with ctapipe : ctapipe_io_nectarcam}. Monte Carlo simulations are briefly presented in section \ref{Monte Carlo simulations} and the Conclusion is done thereafter in section \ref{Conclusion}.

\section[Main Goal : R0 to R1 calibration]{Main Goal : R0 to R1 calibration}
\label{Main Goal : R0 to R1 calibration}

The main goal is to convert the signal in analog-to-digital-conversion ADC to the number of $N_{\mathrm{p.e.}}$ defined by : 
\begin{equation*}
    N_{\mathrm{p.e.}}=\frac{ADC-Ped}{gain} \times FF
\end{equation*}
The pedestal has to be estimated to be substracted to the signal in ADC. Then, the gain has to be calibrated to perform the conversion ADC to p.e., as well as the FF coefficients which are used to balance the inhomogeneities of the camera response.  

Ultimately, the on-site raw data streamed from the CTA camera servers to the CTA Array Control and Data Acquisition system (ACADA), dubbed R1 data in the CTA data model, will include pre-calibrated event data. The pre-calibration requires pedestal subtraction, gain linearisation (conversion from ADC to photo-electron number), gain equalisation among the camera pixels, and the selection of the most appropriate among the two gain channels.

The current version of the camera server software does not perform the pre-calibration required for R1 data. We implemented in \texttt{ctapipe\_io\_nectarcam} (cf. \ref{Interface with ctapipe : ctapipe_io_nectarcam}) a prototype for the pre-calibration software and methods that can read calibration coefficients produced by \texttt{nectarchain} (cf. \ref{nectarchain modules}) and apply them to the raw-data to fill R1 containers. This makes it possible to ingest the data from the camera prototype in the dark room in \texttt{ctapipe}, and thus test the entire analysis chain.\\

\section[nectarchain modules]{\texttt{nectarchain} modules}
\label{nectarchain modules}

\begin{figure}[!h]
\centering
\includegraphics[width=0.8\linewidth]{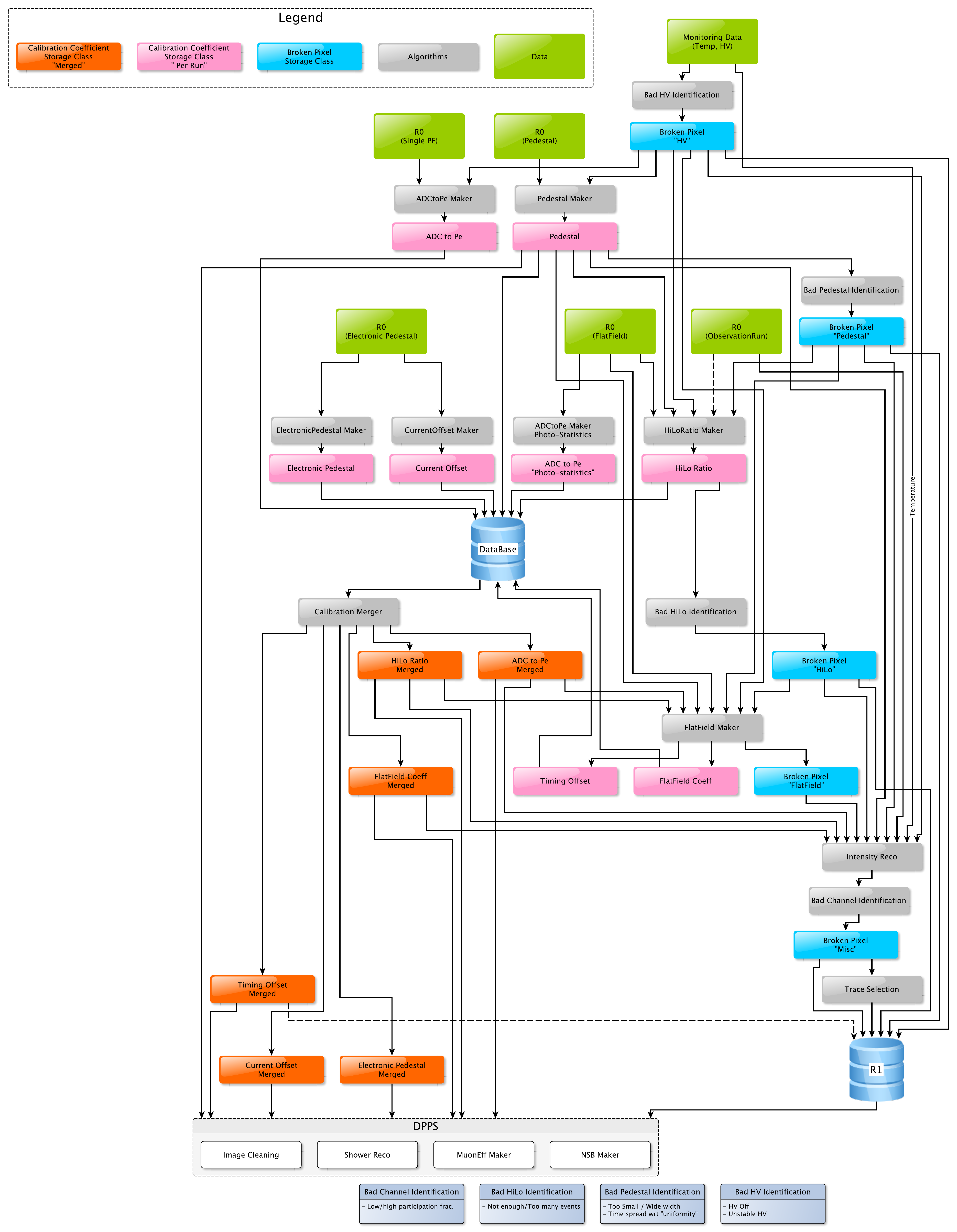}
\caption{Description of the NectarCAM calibration workflow.
\label{fig:module_description}}
\end{figure}

The entire workflow is described in Figure \ref{fig:module_description}. In this part the calibration workflow is presented and the emphasis is made on the main part already implemented.

\subsection[Pedestal estimation]{Pedestal estimation}
\label{Pedestal estimation}

\begin{figure}[!h]
\centering
\includegraphics[width=0.4\linewidth]{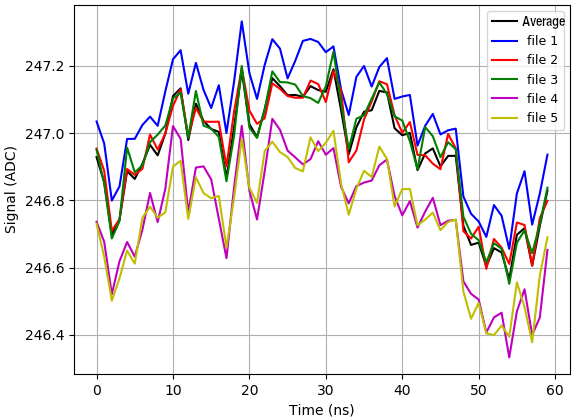}
\caption{Example of pedestal waveforms of a pixel of NectarCAM extracted at 5 different periods of an observation. The waveform labelled "Average" corresponds to the template.
\label{fig:figure_Ped}}
\end{figure}

An accurate pedestal estimate is required to correctly evaluate the charges produced by Cherenkov photons from electromagnetic showers. The pedestal is composed of a deliberately set non-zero offset with electronic noise and, during observations, background light from the night sky. The electronic component can be measured when the camera is in dark conditions (e.g. when the shutter of the camera is closed). For NectarCAM, the average value of the pedestal baseline is set to a value of about 250 ADC. However, it may slightly change with time (warm up of electronics) and with temperature (see Figure \ref{fig:figure_Ped}). Figure \ref{fig:figure_Ped} also shows that the pedestal waveform is affected by fluctuations in the 60~ns window due to slight offsets between the memory cells of the NECTAr chip \citep{2013ICRC...33.3036G} (a part of the NectarCAM front end board which produces ADC conversion). 
During an observation, pedestals are evaluated with periodic pedestal triggers interleaved with regular triggers. A pedestal waveform template is obtained for each pixel and for each channel by averaging waveforms that are not affected by pulses due to noise photons. This template is then used for the analysis of that observation.

\subsection[Flat-field and timing]{Flat-field and timing}
\label{Flat-field and timing}

Additional information needed during the calibration of the data are the relative efficiency between pixels and the relative timing between pixels.
Both are measured with a device called the flat-fielding unit which deliver short pulses of uniform blue light across the camera.
The relative inhomogeneity is obtained by computing the ratio of a given pixel measured intensity to the average across the camera.
The relative timing is obtained by computing the difference between the average maximum time of the waveform to the average across the camera.

\subsection[ADC to photo-electron conversion]{ADC to photo-electron conversion}
\label{ADC to photo-electron conversion}

\subsubsection[Description of the method]{Description of the method}
\label{Description of the method}

The Single photo-electron (SPE) calibration method is based on low illumination calibration data acquisitions where an illuminating device is set to give about 1 p.e. per pixels. Thus, after the accumulation of a reasonable number of events, the charge for each event for each pixel is computed by integrating the ADC signal over a fixed-width window centered on the main pulse. Several charge extraction methods can be used, by varying window width, with a global window for the whole camera or adapted to each pixel for instance. Charge histograms show two main peaks : one created by the pedestal signal, and another one created by the SPE signal. The separation between the two peaks makes it possible to measure the gain, which is the charge in ADC per p.e., with a two Gaussian modelling presented in \cite{Caroff2019}. However, as the illumination is low, this method can only be used to reconstruct the gain in the high gain channel. Derivation of the gain in the low gain channel will then use the High-Low method described in section \ref{H/L ratio}.

\subsubsection[Single photo-electron Calibration devices]{Single photo-electron Calibration devices}
\label{Single photo-electron Calibration devices}

\begin{figure}[!h]
\centering
\includegraphics[width=0.4\linewidth]{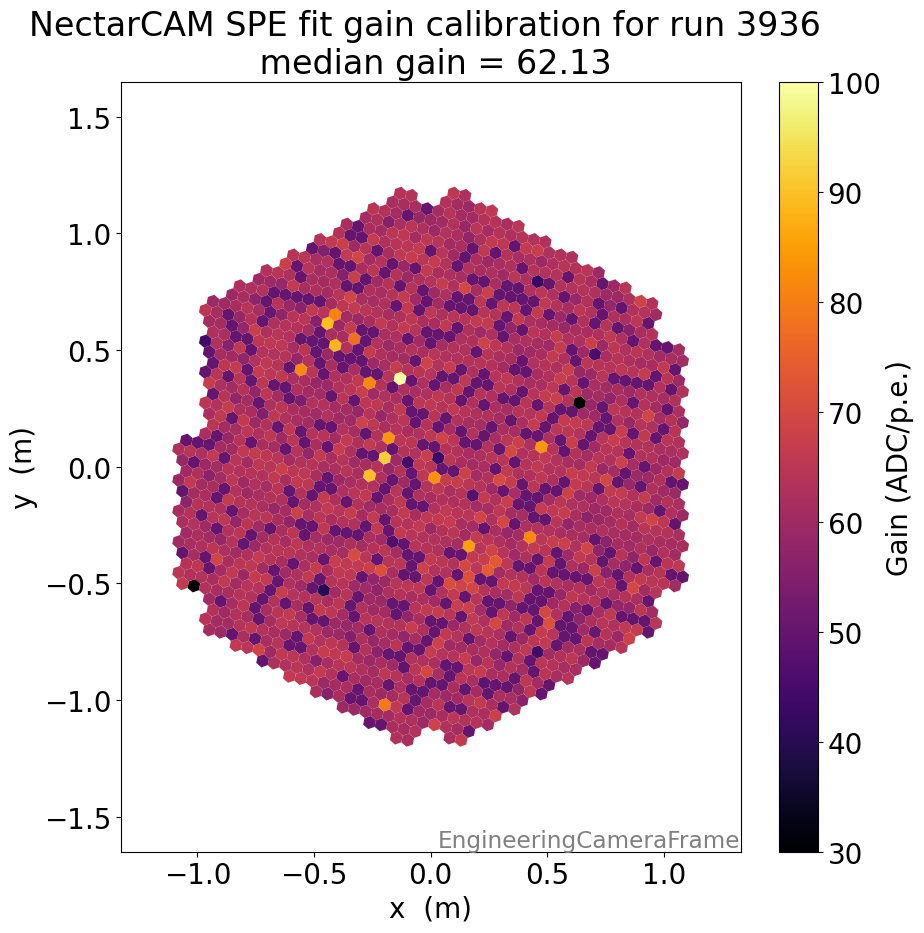}
\caption{Gain computed with the SPE spectrum fit method applied to a FF data acquisition at low illumination.}
\label{fig:SPE_fit_FF}
\end{figure}

Two different types of SPE calibration data acquisitions can be used, the first one is simply a FF one where the flasher is set at low intensity, in this case each pixel is illuminated simultaneously. The gain computed with this method for the entire camera in presented on Figure \ref{fig:SPE_fit_FF}. Moreover, this method can only be performed in the darkroom. On site, the SPE measurement will not be possible with the shutter open (NSB contamination/light pollution) and a special device, embedded in the camera was created to be able to make SPE measurement with the shutter closed. This alternative method does not require simultaneous illumination of the entire camera and is described below.

Analysis of a few of these data acquisitions will be presented in another contribution in these proceedings \citep{2023ICRC_SPE}, as well as the comparison between results from white target runs, FF data acquisitions at low illumination (cf Figure \ref{fig:SPE_fit_FF}), and another method presented in \citep{2023ICRC_SPE} to cross-check all that procedures.

\subsection[High-Low ratio]{High-Low ratio}
\label{H/L ratio}

The measurement of the low gain channel amplification cannot be achieved by direct measurement using e.g the SPE device, as it is expected to be a factor 15 lower in amplification compared to the high gain channel.
However, by using the regime in which both gain channels are linear ($\sim$30-200 p.e.), we are able to measure the ratio of amplification between the high and low gain channels using events that fall in this range.
In NectarCAM, this measurement is done using events from the flat-fielding device set to provide an intensity of $\sim$100~p.e.
The low-gain channel gain value is then the multiplication of the previously calculated high-gain amplification value by this ratio.

\subsection[Data quality monitoring]{Data quality monitoring}
\label{Data quality monitoring}

\begin{figure}[h!]
  \centering
\includegraphics[width=0.40\textwidth, angle =-90]{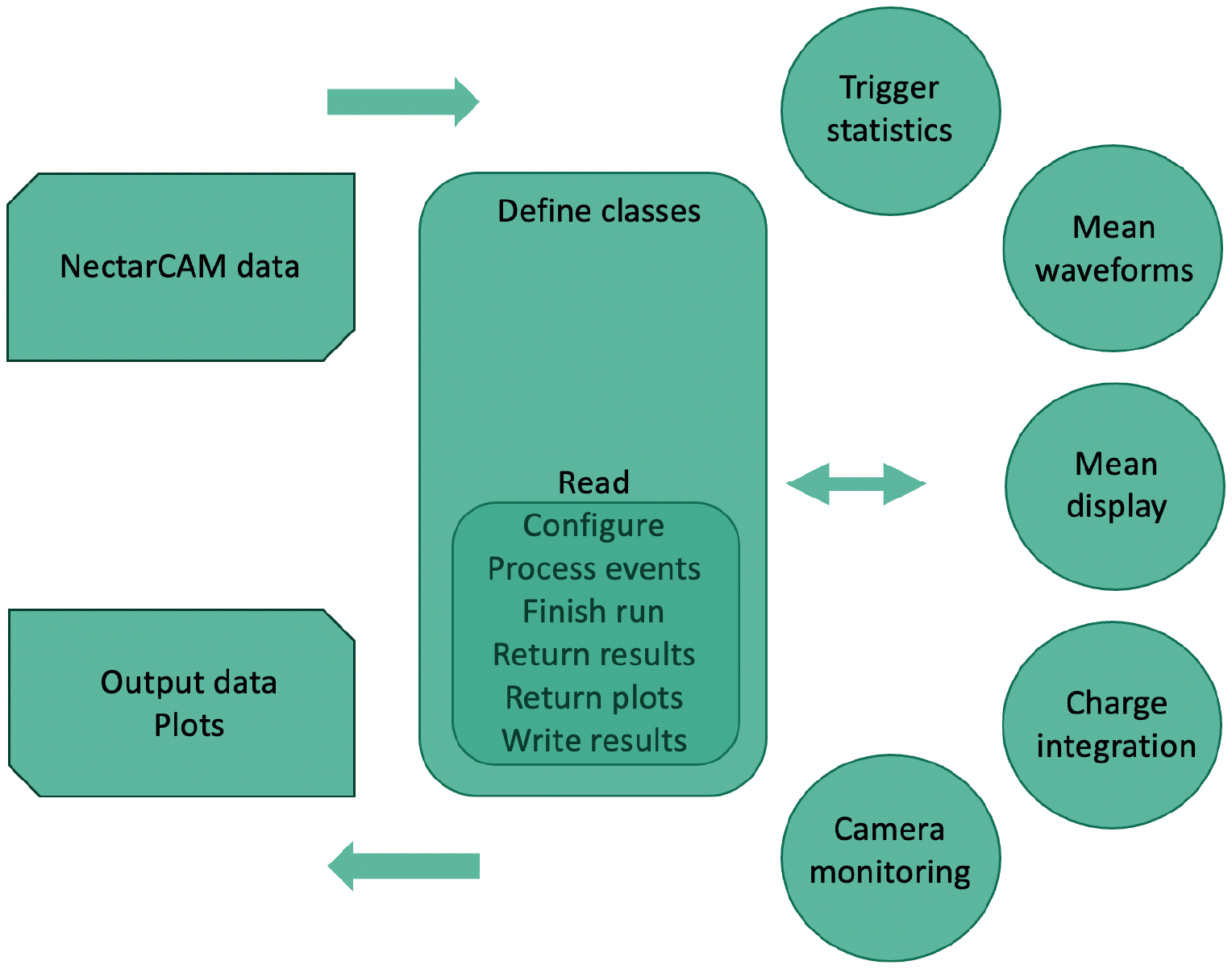}
\caption{DQM semi-automatic pipeline.}
    \label{fig:DQM}
\end{figure}

The data quality monitoring (DQM) pipeline is a semi-automatic pipeline developed for the monitoring of the data taken by the camera and presented in Fig.~\ref{fig:DQM}. It is data acquisition independently of the calibration module and is still in development. It monitors the data for low and high gains and for different types of triggers such as pedestal and physical triggers.   The main script reads the NectarCAM events and defines processor classes that are used to derive the quantities to be monitored. At this stage, the processor contains five classes presented below.

\textbf{Mean waveforms:} The mean waveform class computes the average waveform for all channels and the overall camera average. The waveform of different trigger types can be compared. It allows for the verification of the quality of the camera sampling. 

\textbf{Mean display:} This class averages all the signal seen by the channels through all events in a given data acquisition period. It is used to flag defective pixels, abnormal pedestals and anomalous behavior in general.  

\textbf{Charge integration:} In this class, the charge is integrated around the peak of the signal for each event. An additional option to subtract pedestals, computed from the signal itself, is available. The class computes the charge power spectrum of the camera throughout the entire data acquisition period and computes statistical entities such as the mean, the median, the RMS and the standard deviation of the charge through the whole period. This module can be used to monitor the gain (computation described in section \ref{Single photo-electron Calibration devices}) of the camera.

\textbf{Camera monitoring:} This class monitors the temperature of the modules of the camera. It uses the measurements of the camera by two sensors on the modules saved daily in the log files and matches the temperature with the times of the data acquisition. It also computes the average module temperature during these times. 

\textbf{Trigger statistics:} This class studies the camera trigger rates as a function of trigger ID and time. The triggers are flagged by their nature: pedestal, physical or other.

\subsection[Continuous integration]{Continuous integration}
\label{Continuous integration}

Workflows dedicated to continuous integration (CI) are being developed for the \texttt{nectarchain} package. For instance, the package is published on PyPI on releases, as well as a \texttt{conda} package on the \texttt{conda-forge} channel. Notably, a Singularity/Apptainer \citep{gregory_m_kurtzer_2021_4667718} container image is also built and deployed on the GitHub Container Registry on releases and merged push requests, which greatly eases software deployment. When processing NectarCAM data on the EGI grid through DIRAC \citep{2021EPJWC.25102029A}, where the data are stored, such an image can readily be instantiated without the need of embedding the software in DIRAC jobs or of deploying it to distributed file systems prior to the job processing. Ultimately, \texttt{nectarchain} could be deployed on the CTA Observatory CVMFS\footnote{CernVM File System, \url{https://cvmfs.readthedocs.io/en/stable/}} instance once stabilized. Some work in progress on the CI side concerns well covering unit tests as well as an extensive code documentation.

\section[Interface with ctapipe : ctapipe\_io\_nectarcam]{Interface with \texttt{ctapipe} : \texttt{ctapipe\_io\_nectarcam}}
\label{Interface with ctapipe : ctapipe_io_nectarcam}

The interface between the data acquisition system and the CTA data processing pipeline for NectarCAM is ensured by a dedicated plugin \texttt{ctapipe\_io\_nectarcam}. The plugin reads the data in output from the camera server and fills the containers used by \texttt{ctapipe}, the base library upon which the CTA data processing pipeline is built. The \texttt{ctapipe\_io\_nectarcam} module is distributed through PyPI and as a \texttt{conda} package using the same kind of CI workflows as \texttt{nectarchain}.

\section[Monte Carlo simulations]{Monte Carlo simulations}
\label{Monte Carlo simulations}

To characterize the performance of the NectarCAM camera, in the past we have produced a Monte Carlo model of the camera that was compared to data taken with a partially equipped prototype of the camera at the testbench in CEA Paris-Saclay (France) and on a prototype of the MST structure in Adlershof (Germany) \cite{2022icrc.confE.747A}. As the atmosphere is an integral part of the detection principle, therefore, Monte Carlo simulations are needed to convert the number of photons from the shower to the number of p.e. received in the camera. The NectarCAM camera is now complete and we are updating the Monte Carlo model comparing the simulations to the data taken with the fully equipped camera at the CEA Paris-Saclay testbench taking into account the improved timing resolution analysis \cite{2023arXiv230113828B}.

\section[Conclusion]{Conclusion}
\label{Conclusion}

In these proceedings, the calibration pipeline for NectarCAM has been presented. The most important part is the \texttt{nectarchain} pipeline which is currently used to extract the calibration quatity needed to produce R1 pre-calibrated data. Moreover, \texttt{nectarchain} makes possible study the camera performances (pedestal estimation and evolution in time or with temperature, flat-fielding, gain estimation). \texttt{ctapipe\_io\_nectarcam} is the next part of the calibration workflow and use the output of \texttt{nectarchain} to produce the R1 data in a \texttt{ctapipe} format to be CTAO compliant. The Monte Carlo simulations are still on the rise and will be the next piece of the calibration pipeline. 
The current development status has been presented and we aim to have an operational workflow in the following year.

\acknowledgments

This work was conducted in the context of the CTA Consortium. We gratefully acknowledge financial support from the agencies and organizations listed here:
\url{https://www.cta-observatory.org/consortium_acknowledgments/}.

\bibliographystyle{JHEP}
\bibliography{skeleton.bbl}

%
%
%

\end{document}